\documentclass[a4paper,11pt]{article}

\usepackage{cite}
\usepackage{amsmath}
\usepackage{amscd}
\usepackage{amsfonts}
\usepackage{amssymb}

\begin{document}
\title{\bf{Hamiltonian analysis of symmetries in a massive theory of gravity}}
\author{
{\bf {\normalsize Rabin Banerjee}}\\
 {\normalsize \it S.~N.~Bose National Centre for Basic Sciences,}\\
 {\normalsize \it Block-JD, Sector III, Salt Lake, Kolkata-700098, India.}\\ 
 {\texttt{rabin@bose.res.in}}\\
\and {\bf {\normalsize Sunandan Gangopadhyay}}{\footnote{Also, Visiting Associate at S.~N.~Bose National Centre for Basic Sciences, JD Block, Sector III, Salt Lake, Kolkata-700098, India.}}\\
 {\normalsize \it Department of Physics,}\\
 {\normalsize \it West Bengal State University,}\\
 {\normalsize \it Barasat, North 24 Paraganas, West Bengal, India.}\\
  {\texttt{sunandan.gangopadhyay@gmail.com} \& \texttt{sunandan@bose.res.in}}\\
\and {\bf {\normalsize Debraj Roy}}\\
 {\normalsize \it S.~N.~Bose National Centre for Basic Sciences,}\\
 {\normalsize \it Block-JD, Sector III, Salt Lake, Kolkata-700098, India.}\\
 {\texttt{debraj@bose.res.in}}
}
\vspace{-0.5truecm}

\date{}

\maketitle

\begin{abstract}
We construct the generator of hamiltonian gauge symmetries in a 2+1 dimensional massive theory of gravity, proposed recently, through a systematic off-shell algorithm. Using a field dependent map among gauge parameters we show that the symmetries obtained from this generator are on-shell equivalent to the Poincar\'{e} gauge symmetries. We also clarify certain subtle issues concerning the implementation of this map.
\end{abstract}

\section{Introduction}
\label{Sec:Intro}

A unitary, renormalizable theory of gravity with propagating degree(s) of freedom is a long sought goal towards our understanding of gravitation. Recently, such a proposal (`new massive gravity' or `BHT gravity' \cite{Bergshoeff:2009hq, Bergshoeff:2009aq}) with massive propagating modes in 3 spacetime dimensions has generated much interest \cite{Deser:2009hb,Oda:2009ys,Clement:2009gq,Gullu:2010sd,Gullu:2010pc,Blagojevic:2010ir,Blagojevic:2011qc,Perez:2011qp,Jatkar:2011ue,Ahmedov:2011}, having particular emphasis on its symmetries \cite{Blagojevic:2010ir,Blagojevic:2011qc}. A massive spin 2 description is quite standard. At the linearised level of Einstein gravity, we have the standard non-interacting Fierz-Pauli (FP) model \cite{Fierz:1939ix} in any dimension. It is unitary, and in 3D has two massive degrees of freedom. Also in 3D, addition of a Chern-Simons term (in the connection variables) gives the topologically massive gravity (TMG) model \cite{Deser:1982vy,Deser:1981wh}. This theory violates parity and has one propagating degree of freedom. The BHT gravity is unitary and can give an interacting theory at the non-linear level, unlike the FP model. It is given by the action
\begin{align}
\label{action BHT simple}
S = \frac{1}{\kappa^2} \int d^3x ~\sqrt{g}\left[ R + \frac{1}{m^2} K \right],
\end{align}
where $R$, the Ricci scalar is the contraction of the Ricci tensor ($R^\mu_{\ \mu}$) and $K=R_{\mu\nu}R^{\mu\nu} - \tfrac{3}{8}R^2$. The BHT model can be both motivated as a non-linear generalization of the FP model, or a soldering of two TMG massive modes \cite{Dalmazi:2009es,Bergshoeff:2009fj}. The interesting point to note is that the model is unitary in spite of fourth order derivatives present in the action, through the $K$ term. 

Now, interest in 3D gravity is fuelled by studies on the  the AdS/CFT correspondence and towards understanding fundamental problems such as entropy of the (BTZ-type) black-hole solutions. The BHT action \eqref{action BHT simple} can incorporate a cosmological term to give the action
\begin{align}
\label{action BHT cosmo}
S = \frac{1}{\kappa^2} \int d^3x ~\sqrt{g}\left[\sigma R + \frac{1}{m^2} K - 2 \lambda m^2 \right].
\end{align}
The unitarity and stability of this model depends in general on the choice of parameters and unitarity has been studied in different regions of the parameter space (see for example, \cite{Bergshoeff:2009aq}).

A consistent canonical constraint analysis of BHT gravity has been carried out in \cite{Sadegh:2010pq,Blagojevic:2010ir}. In \cite{Blagojevic:2010ir}, this was done in the first-order formulation through the Poincar\'{e} gauge theory (PGT) construction \cite{Utiyama:1956sy, Kibble:1961ba,Sciama:1962,Hehl:1976kj,Blagojevic:2002du}. However the gauge generator (from which transformation of the basic fields are obtained) is constructed by an on-shell algorithm due to Castellani \cite{Castellani:1981us}. This view of symmetries is restricted in the sense that it views symmetries as maps between solutions to solutions of the equations of motion, rather than as a map between field configurations.

In this paper, we shall construct the gauge generator of cosmological BHT gravity \cite{Bergshoeff:2009aq} through a hamiltonian algorithm following \cite{Banerjee:1999yc,Banerjee:1999hu} which is off-shell and uses on the total hamiltonian.\footnote{See \cite{Henneaux:1990au} for a treatment of symmetries using the extended hamiltonian formulation.} This procedure has been used recently in the context of diffeomorphism symmetry in string theory \cite{Banerjee:2004un,Banerjee:2005bb}, second order metric gravity \cite{Mukherjee:2007yi}, interpolating formulation of bosonic string theory \cite{Gangopadhyay:2007gn}, and also in topological gravity with torsion in (2+1)-dimensions \cite{Banerjee:2009vf}. The formalism treats gauge symmetries of the action at an off-shell level, i.e. it views symmetries as maps between field configurations in the action.

The derivation of the generator in the present case of BHT gravity is more subtle. The theory contains second-class constraints which have not been removed completely through Dirac brackets as is usually done (see, for instance, \cite{Banerjee:2009vf}). This is in complete contrast to earlier examples where the theories either comprised of first-class constraints and/or second-class constraints which were totally removed through Dirac brackets.

After constructing the generator, we will give explicit expression of the symmetries of the basic fields. It will be shown that these symmetries can be mapped to the underlying Poincar\'{e} symmetries through a field dependent map between gauge parameters. This mapping is only possible \emph{on-shell}, i.e. upon imposition of the equations of motion. In particular, we show that the symmetry of the triad field $b^i_{\ \mu}$, which is related to the metric through $g_{\mu\nu}=b^i_{\ \mu}b^j_{\ \nu}\,\eta_{ij}$, is identifiable with the Poincar\'{e} symmetries upon imposition of an equation of motion that implies zero torsion. This is interesting, as it is precisely the condition of zero torsion that takes us from the Riemann-Cartan spacetime of PGT to the Riemannian spacetime in which BHT was originally formulated, adopting a usual metric formalism. We will also show that though this map is field dependent, it can be used in the generator both after and before computing symmetries, equivalently, to relate the two sets of symmetries.

We would like to emphasize that the computation of the canonical generator at an off-shell level would be important in cleanly obtaining the conserved charges of BHT gravity. Since black hole solutions are known to exist in this theory \cite{Clement:2009gq,Clement:2009ka,Oliva:2009ip,Myung:2011bn}, hence a knowledge of the conserved charges would prove to be useful in studying various thermodynamic aspects, like entropy and area law.

Let us now briefly explain the organisation of our article. We begin in section 2 with the canonical description of the model following \cite{Blagojevic:2010ir}, listing all the first-class and second-class constraints of the theory. In section 3 we systematically construct the off-shell generator of gauge symmetries in detail. In the following section 4, we study the hamiltonian gauge symmetries of the basic fields by employing the generator constructed in the previous section. We also study the relation of these symmetries to the Poincar\'{e} symmetries through a mapping of the gauge parameters. In this same section, we demonstrate a mechanism for a consistency check of our algorithm. Finally, we conclude in section 4 with a short summary. And below, we give a description of some important notational conventions adopted in our calculations.

\noindent{\em Summary of conventions:} Latin indices refer to the local Lorentz frame and the Greek indices refer to the coordinate frame. The beginning letters of both alphabets $(a,b,c,\ldots)$ and $(\alpha,\beta,\gamma,\ldots)$ run over the space part (1,2) while the middle alphabet letters $(i,j,k,\ldots)$ and $(\mu,\nu,\lambda,\ldots)$ run over all coordinates (0,1,2). The totally antisymmetric tensor $\varepsilon^{ijk}$ and the tensor density $\varepsilon^{\mu\nu\rho}$ are both normalized so that $\varepsilon^{012}=1$. The signature of space-time adopted here is $\eta=(+,-,-)$.

\section{Canonical description of the model}
\label{Sec:Model}

We begin our analysis with first order form of BHT massive gravity, written in accordance with the PGT formalism, where the basic variables are the triads $b^i_{\ \mu}$ and spin connections $\omega^i_{\ \mu}$ \cite{Blagojevic:2010ir}. The formulation of PGT starts on a globally flat space (here $3D$) with a local set of orthogonal coordinates at each point. Any global field $A^\mu$ is written in terms of these local coordinates $A^i$ by a set of vielbein fields `$b$' (triads) as $A^\mu(x) = b^i_{\ \mu}(x) A_i(x)$. The Lagrangian is made invariant under global Poincar\'{e} transformations by construction. Localisation of this global Poincar\'{e} symmetry demands the introduction of covariant derivatives $\nabla_\mu = \partial_\mu + \texttt{Conn}_\mu$ through compensating connection variables `$\texttt{Conn}$' in the standard manner of constructing gauge theories. The respective field strengths defined through the commutator of the covariant derivatives gives the Riemann tensor $R^i_{\ \mu\nu}$ and the torsion $T^i_{\ \mu\nu}$:
\begin{align}
\label{PGT R T}
\begin{aligned}
R^i_{\mu\nu} &= \partial_\mu \omega^i_{\ \nu} - \partial_\mu \omega^i_\nu + \epsilon^i_{\ jk}\omega^j_{\ \mu}\omega^k_{\ \nu} \\
T^i_{\ \mu\nu} &= \nabla_\mu b^i_{\ \nu} - \nabla_\nu b^i_{\ \mu}.
\end{aligned}
\end{align}
Here the covariant derivative of the triad is defined as $\nabla_\mu  b^i_{\ \nu} = \partial_\mu b^i_{\ \nu} + \epsilon^i_{\ jk}\omega^j_{\ \mu}b^k_{\ \nu}$, with $\omega^j_{\ \mu}$ being the `spin connections' arising out of the connection part $\texttt{Conn}$ of the covariant derivative. The spacetime naturally occurring in this construction is thus the {\em Riemann-Cartan} spacetime with non-zero torsion. The transformation of the basic fields under the Poincar\'{e} transformations are:
\begin{align}
\label{PGT deltas}
\begin{aligned}
\delta_{\scriptscriptstyle PGT} b^i_{\ \mu} &= -\epsilon^i_{\ jk}b^j_{\ \mu}\theta^k - \partial_\mu \xi^\rho \,b^i_{\ \rho} - \xi^\rho\,\partial_\rho b^i_{\ \mu} \\
\delta_{\scriptscriptstyle PGT} \omega^i_{\ \mu} &= -\partial_\mu \theta^i - \epsilon^i_{\ jk}\omega^j_{\ \mu}\theta^k - \partial_\mu\xi^\rho\,\omega^i_{\ \rho} - \xi^\rho\,\partial_\rho\omega^i_{\ \mu}.
\end{aligned}
\end{align}
In the above symmetries, the parameter describing local Lorentz transformations is $\theta^i(x)$ and that describing general coordinate transformations is $\xi^\mu$, both transformations being of infinitesimal order. It is to be noted that the nature of these transformations depend on the behaviour of a field under the action of the Lorentz group. Thus any field having the general nature of the triad field $b^i_{\ \mu}$, i.e. which transforms as a vector in both spaces, have the same transformations as given above in \eqref{PGT deltas}. In particular, we list the transformations of two fields `$\lambda$' and `$f$' which will be required later in this article,
\begin{align}
\label{PGT deltas lambda f}
\begin{aligned}
\delta_{\scriptscriptstyle PGT} \lambda^i_{\ \mu} &= -\epsilon^i_{\ jk}\lambda^j_{\ \mu}\theta^k - \partial_\mu \xi^\rho \,\lambda^i_{\ \rho} - \xi^\rho\,\partial_\rho \lambda^i_{\ \mu} \\
\delta_{\scriptscriptstyle PGT} f^i_{\ \mu} &= -\epsilon^i_{\ jk}f^j_{\ \mu}\theta^k - \partial_\mu \xi^\rho \,f^i_{\ \rho} - \xi^\rho\,\partial_\rho f^i_{\ \mu}.
\end{aligned}
\end{align}

The first-order BHT model we work with, to begin with, contains the usual Einstein-Hilbert piece along with a cosmological term. Now, PGT is formulated on the Riemann-Cartan spacetime where both curvature and torsion play their parts. Originally however, BHT gravity was formulated in the Riemann spacetime, which has zero torsion. To be able to enforce this condition, torsion is included in the action via coupling to a dynamical Lagrange multiplier field $\lambda^i_{\ \mu}$. The distinctive term of the BHT theory which contains the square of curvature is incorporated into the action with the help of an auxiliary field, such that the action is rendered linear in curvature. On imposition of the equation of motion for $f^i_{\ \mu}$, the curvature square term of original BHT is recovered. The lagrangian, with all the above described terms and their individual coupling parameters, take the following form:
\begin{align}
\label{lagrangian}
\mathcal{L}=a\epsilon^{\mu\nu\rho}\left( \sigma b^i_{\ \mu}R_{i\nu\rho}-\frac{\Lambda}{3}\epsilon_{ijk}b^i_{\ \mu}b^j_{\ \nu}b^k_{\ \rho} \right) + \frac{a}{m^2} \mathcal{L}_K + \frac{1}{2} \epsilon^{\mu\nu\rho}\lambda^i_{\ \mu}T_{i\nu\rho}.
\end{align}
Here $R_{i\nu\rho}$ and $T_{i\nu\rho}$ are the Riemann tensor and torsion defined earlier, while $\mathcal{L}_K$ is defined as:
\begin{align}
\label{L_K}
\mathcal{L}_K&=\frac{1}{2} \epsilon^{\mu\nu\rho}f^i_{\ \mu}R_{i\nu\rho}-b\mathcal{V}_K\nonumber\\
\mathcal{V}_K&=\frac{1}{4}\left(f_{i\mu}f^{i\mu}-f^2\right),
\end{align}
where $b$ denotes the determinant of the basic triad field $b^i_{\ \mu}$. The equations of motion corresponding to variations with respect to the basic variables $b^i_{\ \mu}$, $\omega^i_{\ \mu}$, $f^i_{\ \mu}$ and $\lambda^i_{\ \mu}$, respectively, are given below:
\begin{subequations}
\label{EOM}
\begin{align}
\label{EOM b}
& a \epsilon^{\mu \nu \rho} \left( \sigma R_{i\nu\rho} - \Lambda \epsilon_{ijk}b^j_{\ \nu}b^j_{\ \rho} \right) - \frac{ab}{m^2} \,\mathcal{T}_i^{\ \mu} + \epsilon^{\mu \nu \rho} \nabla_\nu \lambda_{i\rho} - \frac{ab}{2m^2}\,\Theta_{ij}\left( f^{j\mu}-b^{j\mu} \right) = 0\\
\label{EOM omega}
& \epsilon^{\mu\nu\rho} \left[ a\sigma T_{i\nu\rho} + \frac{a}{m^2} \nabla_\nu f_{i\rho} + \epsilon_{ijk} b^j_{\ \nu} \lambda^k_{\ \rho} \right] = 0\\
\label{EOM f}
& \frac{a}{2m^2} \left[ \epsilon^{\mu\nu\rho}R_{i\nu\rho} - b\left( f_i^{\ \mu} - f b_i^{\ \mu} \right) \right] = 0 \\
\label{EOM lambda}
& \frac{1}{2} \,\epsilon^{\mu\nu\rho} \, T_{i\nu\rho} = 0.
\end{align}
\end{subequations}
Here $f=f^h_{\ \rho}\,b_h^{\ \rho}$ is the trace of the field $f^i_{\ \mu}$ and $\mathcal{T}_i^{\ \mu}$ is defined as:
\begin{align}
\mathcal{T}_i^{\ \mu} = b_i^{\ \mu} \mathcal{V}_K - \frac{1}{2} \left( f_{ik}f^{k\mu} - f f_i^{\ \mu} \right).
\end{align}
The term $\Theta_{ij} = f_{ij} - f_{ji}$ is proportional to the antisymmetric part of the field $f^i_{\ \mu}$. Similarly, we can define an antisymmetric combination from $\lambda^i_{\ \mu}$ as $\Psi_{ij}=\lambda_{ij}-\lambda_{ji}$. The equations of motion however show that both fields $f_{ij}$ and $\lambda_{ij}$ are symmetric \cite{Blagojevic:2010ir}. Hence $\Theta_{ij} = \Psi_{ij} = 0$. Later, in this section itself, we see that $\Theta_{ij}$ and $\Psi_{ij}$ appear as constraints of the theory. Thus the symmetry of the auxiliary fields is also a result of the constraint structure and does not involve a true equation of motion (involving accelerations).

Next, we summarize the hamiltonian description of the theory along with a proper identification of the constraints {\em \`{a} la} Dirac, following \cite{Blagojevic:2010ir}. The momenta corresponding to the basic fields are defined in the standard manner $p=\frac{\partial \mathcal{L}}{\partial \dot{q}}$.
\begin{table}[h]
\label{Tab:Momenta}
\centering
\begin{tabular}{l|cccc}
\hline\hline
Basic Field & $b^i_{\ \mu}$ & $\omega^i_{\ \mu}$ & $f^i_{\ \mu}$ & $\lambda^i_{\ \mu}$ \\[0.5ex] \hline
Conjugate Momenta & $\pi_i^{\ \mu}$ & $\Pi_i^{\ \mu}$ & $P_i^{\ \mu}$ & $p_i^{\ \mu}$ \\[0.5ex]\hline\hline
\end{tabular}
\caption{The basic fields and their corresponding momenta}
\end{table}
The canonical hamiltonian, defined as $\mathcal{H}_C=p\dot{q}-\mathcal{L}$, after some rearrangements is given by:
\begin{align}
\label{H_C}
\mathcal{H}_C = b^i_{\ 0}\hat{\mathcal{H}}_i + \omega^i_{\ 0}\mathcal{K}_i + f^i_{\ 0}\hat{\mathcal{R}}_i + \lambda^i_{\ 0}\mathcal{T}_i + \partial_\alpha \mathcal{D}^\alpha,
\end{align}
where the relevant quantities are defined below:-
\begin{align}
\label{H_C defns}
\begin{aligned}
\hat{\mathcal{H}}_i &= \mathcal{H}_i + \frac{a}{m^2}\: b \: \mathcal{T}_i^{\ 0} \\
\mathcal{H}_i &= \epsilon^{0\alpha\beta} \left( a\sigma R_{i\alpha\beta} - a\Lambda\:\epsilon_{ijk}b^j_{\ \alpha}b^k_{\ \beta} + \nabla_\alpha \lambda_{i\beta} \right)\\
\mathcal{K}_i &= -\epsilon^{0\alpha\beta} \left( a\sigma T_{i\alpha\beta} + \frac{a}{m^2}\:\nabla_\alpha f_{i\beta} + \epsilon_{ijk} b^j_{\ \alpha} \lambda^k_{\ \beta} \right)\\
\mathcal{R}_i &= -\frac{a}{2m^2}\,\epsilon^{0\alpha\beta}\,R_{i\alpha\beta}\\
\hat{\mathcal{R}}_i &= \mathcal{R}_i + \frac{a}{2m^2} b \left( f_i^{\ 0} - f b_i^{\ 0} \right) \\
\mathcal{T}_i &= -\frac{1}{2}\:\epsilon^{0\alpha\beta}\,T_{i\alpha\beta} \\
\mathcal{D}^\alpha &= \epsilon^{0\alpha\beta} \left[ \omega^i_{\ 0}\left( 2a\sigma b_{i\beta} + \frac{a}{m^2}f_{i\beta} \right) + b^i_{\ 0}\lambda_{i\beta} \right].
\end{aligned}
\end{align}

The canonical analysis of this model, done in \cite{Blagojevic:2010ir}, treats the second-class sector in a mixed manner. It employs a set of Dirac brackets to eliminate a sector of the second-class constraints that arise from the primary sector, and fixes the Lagrange multipliers corresponding to the other set. Now, it turns out that all the momenta give rise to primary constraints. These are listed below:
\begin{subequations}
\label{primary}
\begin{align}
\phi_i^{\ \mu} &:= \pi_i^{\ \mu} - \epsilon^{0\alpha\beta} \lambda_{i\beta}\;\delta_\alpha^\mu \approx 0\\
\Phi_i^{\ \mu} &:= \Pi_i^{\ \mu} - 2a\epsilon^{0\alpha\beta} \left( \sigma b_{i\beta} + \frac{1}{2m^2}f_{i\beta} \right) \: \delta_\alpha^\mu \approx 0 \\
P_i^{\ \mu} &\approx 0 \; ; \qquad\qquad p_i^{\ \mu} \approx 0.
\end{align}
\end{subequations}
Among the above primary constraints, the set $\,X := \left( \pi_i^{\ \alpha}, \Pi_i^{\ \alpha}, P_i^{\ \alpha}, p_i^{\ \alpha} \right)$ fixes the corresponding Lagrange multipliers in an appropriately defined total hamiltonian $$\mathcal{H}_T = \mathcal{H}_C + \text{sum of all primary constraints,}$$ and hence are second class in nature. To eliminate this sector $X$ an appropriate set of Dirac brackets is introduced. Consequently, the momenta $\,\left( \pi_i^{\ \alpha}, \Pi_i^{\ \alpha}, P_i^{\ \alpha}, p_i^{\ \alpha} \right)$ can now be eliminated and the analysis is carried in a {\em reduced phase space} with a modified algebra, given below:
\begin{subequations}
\label{algebra1}
\begin{align}
\lbrace b^i_{\ \alpha}, \lambda^j_{\ \beta}\rbrace^* &= \eta^{ij}\epsilon_{0\alpha\beta}\\
\lbrace \omega^i_{\ \alpha}, f^j_{\ \beta} \rbrace^* &= \left( \frac{m^2}{a} \right) \eta^{ij} \epsilon_{0\alpha\beta}\\
\lbrace \lambda^i_{\ \alpha}, f^j_{\ \beta} \rbrace^* &= \left( -2m^2\sigma \right) \eta^{ij} \epsilon_{0\alpha\beta}.
\end{align}
\end{subequations}
The other brackets in this new algebra turn out to be same as the corresponding Poisson brackets. In particular, we note the following brackets, derivable using \eqref{algebra1} and using the inverse property of the triad field $b^i_{\ \mu}b_j^{\ \mu}=\delta^i_j$,
\begin{align}
\label{algebra2}
\begin{aligned}
\lbrace b_i^{\ \mu}, \pi_j^{\ \nu} \rbrace^* &= b_j^{\ \mu}b_i^{\ \nu}\\
\lbrace b_i^{\ \mu}, \lambda^j_{\ \nu} \rbrace^* &= \epsilon_{0\alpha\beta} \, b_i^{\ \beta}b^{j\mu}\:\delta^\alpha_\nu.
\end{aligned}
\end{align}
Since no Poisson brackets are employed in our analysis and {\em all} our brackets correspond to this reduced space algebra, we will henceforth drop the starred bracket notation and indicate the changed algebra with usual braces, i.e. $\lbrace,\rbrace^*:=\lbrace,\rbrace$.

The final constraint structure in our reduced space is presented in Table \ref{Tab:Constraints}
\begin{table}[h]
\label{Tab:Constraints}
\centering
\begin{tabular}{l c c}
\hline\hline\
& First Class & Second class \\[0.2ex] \hline\\[-1.9ex]
Primary & $\Sigma_{(3) i} = \pi''{}_i^{\ 0}\;, \;\;\Sigma_{(4) i} = \Pi_i^{\ 0}$ & $p_i^{\ 0}$, $P_i^{\ 0}$ \\[0.4ex]
Secondary & $\Sigma_{(1) i} = \bar{\mathcal{H}}_i\;, \;\;\Sigma_{(2) i} = \bar{\mathcal{K}}_i$ & $\mathcal{T}_i$, $\hat{\mathcal{R}'}_i$ \\[0.4ex]
Tertiary & & $\Theta_{ij}$, $\Psi_{ij}$ \\[0.4ex]
Quartic & & $\chi$, $\varphi$ \\[0.4ex]
\hline\hline
\end{tabular}
\caption{Constraints under the modified algebra classified.}
\end{table}
and the required quantities are defined below:
\begin{align}
\label{Constrnt defns}
\begin{aligned}
\pi''{}_i^{\ 0} &:= \pi_i^{\ 0} + f_i^{\ l}P_l^{\ 0} + \lambda^l_{\ i}\,p_l^{\ 0}\\
\bar{\mathcal{H}}_i &:= \hat{\mathcal{H}}_i + f^l_{\ i}\hat{\mathcal{R}}_l + \lambda^l_{\ i}\mathcal{T}_l + b_i^{\ \rho}(\nabla_\rho \lambda_{jk})b^k_{\ 0}\,p^{j0} + b_i^{\ \rho}(\nabla_\rho f_{jk})b^k_{\ 0}\,P^{j0}\\
\bar{\mathcal{K}}_i &:= \mathcal{K}_i - \epsilon_{ijk}\left( \lambda^j_{\ 0}\,p^{k0} - b^j_{\ 0}\lambda^k_{\ l}\,p^{l0} \right) - \epsilon_{ijk}\left( f^j_{\ 0}\,P^{k0} - b^j_{\ 0} f^k_{\ l}\,P^{l0} \right)\\
\varphi &:= \sigma f + 3\Lambda_0 + \frac{1}{2m^2}\:\mathcal{V}_K\\
\chi &:= \lambda^i_{\ \mu}b_i^{\ \mu} = \lambda.
\end{aligned}
\end{align}

The total hamiltonian density in the reduced phase space may be defined at first as the canonical hamiltonian plus {\em all} primary constraints that have not been eliminated, i.e.
\begin{align}
\label{H_T 2}
\mathcal{H}_T &= \mathcal{H}_C + u^i_{\ 0}\phi_i^{\ 0} + v^i_{\ 0} \Phi_i^{\ 0} + w^i_{\ 0}p_i^{\ 0} + z^i_{\ 0}P_i^{\ 0}.
\end{align}
However, in the reduced phase space, $p_i^{\ 0}$ and $P_i^{\ 0}$ are both second-class and are consequently used to fix the multipliers $w^i_{\ 0}$ and $z^i_{\ 0}$. These can now be added to the canonical hamiltonian density $\mathcal{H}_C$ to form a new quantity, often denoted as $\mathcal{H}^{(1)}$
\begin{align}
\label{H1 defn}
\mathcal{H}^{(1)} &:= \mathcal{H}_C + \text{sum of primary second-class constraints with determined multipliers}\nonumber\\
&= b^i_{\ 0}\bar{\mathcal{H}}_i + \omega^i_{\ 0}\bar{\mathcal{K}}_i
\end{align}
The total hamiltonian density now becomes:
\begin{align}
\label{H_T 3}
\mathcal{H}_T &= \mathcal{H}^{(1)} + \text{sum of all primary first-class constraints with arbitrary multipliers}\nonumber\\
&= b^i_{\ 0}\bar{\mathcal{H}}_i + \omega^i_{\ 0}\bar{\mathcal{K}}_i + u^i_{\ 0}\pi''{}_i^{\ 0} + v^i_{\ 0}\Pi_i^{\ 0}.
\end{align}
In the next section we see, that, it is this modified hamiltonian density $\mathcal{H}^{(1)}$ which becomes useful in our construction of symmetry generators of this mixed-model, with both first and second-class sectors. It plays a part analogous to that played by $\mathcal{H}_C$ in systems with only first-class sector or systems where the second-class sector is {\em completely} eliminated using Dirac brackets.

\section{Construction of the hamiltonian generator}
\label{Sec:Gen}

In this section we proceed to systematically construct an off-shell generator of the model \eqref{lagrangian} following the method shown in \cite{Banerjee:1999yc, Banerjee:1999hu}. Let us denote the relevant (first-class) constraints in our theory (see Table \ref{Tab:Constraints}) as:
\begin{align}
\label{RB const1}
\Sigma_{(I)} = \left[\Sigma_{(A)};\Sigma_{(Z)}\right],
\end{align}
where $A=3,4$ are primary (first class) constraints, $Z=1,2$ secondary (first class) constraints and $I=1,2,3,4$ constitute all (first class) constraints. The total hamiltonian density \eqref{H_T 3} may then be written as
\begin{align}
\label{H_T chi}
\mathcal{H}_T = \mathcal{H}^{(1)} + \chi^{(A)} ~\Sigma_{(A)},
\end{align}
with the notation $\chi^{(3)}=u^i_{\ 0}$ and $\chi^{(4)}=v^i_{\ 0}$.

By a gauge generator we mean a field dependent quantity $G$, such that for any quantity $F$ which is a function of the basic fields, the bracket $\lbrace F, G \rbrace$ gives the variation $\delta F$ consistent with the variations of the basic fields. In particular we then have
\begin{align}
\label{G q defn}
\delta q = \lbrace q, G \rbrace.
\end{align}
Now, the Dirac prescription for the generator is to consider a linear combination of {\em all} first class constraints
\begin{align}
\label{gen G}
G = \int d^2x ~\varepsilon^{(I)}\Sigma_{(I)}
\end{align}
where $\varepsilon^{(I)}$ are the gauge parameters. However, not all of these are independent. We have to now eliminate the dependent parameters and write the generator in terms of the independent gauge parameters alone.

We start by noting that the gauge variations are not completely arbitrary, but must commute with time derivatives i.e.
\begin{align}
\label{dotdelta}
\left( \delta \bullet \frac{d}{dt} \right) q \equiv \left( \frac{d}{dt} \bullet \delta \right) q,
\end{align}
where $\dfrac{d}{dt} q = \lbrace q, \int \mathcal{H}_T\rbrace$. Both sides of \eqref{dotdelta} can be evaluated separately using the generator \eqref{gen G} and the total hamiltonian \eqref{H_T 3}. The generator is composed of the first class constraints and the total hamiltonian density is the sum of $\mathcal{H}^{(1)}$ and the primary first class constraints. So the algebrae required will be those in-between the first class constraints and that of the first class sector with $\mathcal{H}^{(1)}$. For calculation, we introduce some structure functions and calculate these required algebrae.

By a theorem due to Dirac \cite{Dirac:Lectures} the first-class constraints must close amongst themselves, i.e.
\begin{align}
\label{C_defn}
\left\lbrace\Sigma_{(I) i}(x),\Sigma_{(J) j}(x')\right\rbrace = &\int d^2x''\,{\left(C^K_{\;\;\, IJ}\right)}_{ijk}(x'',x,x') ~\Sigma_{(K)}^{\quad k}(x'').
\end{align}
Also note that $\mathcal{H}^{(1)}$ \eqref{H1 defn} is a first-class quantity as the total hamiltonian must be first-class. This is analogous to the first-class nature of the canonical hamiltonian in a system with only first-class constraints. So we must have
\begin{align}
\label{V_defn}
\left\lbrace \int \mathcal{H}^{(1)},\Sigma_{(I)i}(x)\right\rbrace = &\int d^2x'\,{\left(V^J_{\;\;\: I}\right)}_{ik}(x',x)~\Sigma_{(J)}^{\quad k}(x').
\end{align}
Using the above definitions \eqref{C_defn} and \eqref{V_defn} in \eqref{dotdelta}, we reach the following set of equations relating the gauge parameters \cite{Banerjee:1999hu, Banerjee:1999yc}:
\begin{align}
\label{RB master 1}
\delta\chi^{(A)}(x) &= \displaystyle\frac{d\varepsilon^{(A)}(x)}{dt} - \int d^2x' ~\varepsilon^{(I)}(x') \,\left[ \left(V^A_{\ I}\right)(x,x')+\int d^2x''\,\chi^{(B)}(x'') ~\left(C^A_{\ IB}\right)(x,x',x'')\right]\\
\label{RB master 2}
0 &= \displaystyle\frac{d\varepsilon^{(Z)}(x)}{dt} - \int d^2x' ~\varepsilon^{(I)}(x') \left[ \left(V^Z_{\ I}\right)(x,x') +\int d^2x''\,\chi^{(B)}(x'') \,\left(C^Z_{\ IB}\right)(x,x',x'')\right].
\end{align}
Among them, the second condition makes it possible to choose $(A)$ independent gauge parameters from the set $\varepsilon^{(I)}$ and express the generator $G$ (\ref{gen G}) entirely in terms of them. This shows that the number of independent gauge parameters is equal to the number of independent, primary first-class constraints \cite{Gomis:1989vy}. As for the first condition, it does not impose any new condition on the gauge parameters $\varepsilon$. It is actually a consistency check of the whole scheme as it can be independently derived, using the second equation and the generator constructed \cite{Banerjee:1999yc,Banerjee:1999hu}. We will demonstrate this explicitly in the case of our model, later.

Note that the derivation of (\ref{RB master 2}) is based only on the relation between the velocities and the canonical momenta, namely, the first of the Hamilton's equations of motion \cite{Banerjee:1999hu, Banerjee:1999yc}. The full dynamics, implemented through the second of Hamilton's equations $\left(\frac{dp}{dt}=\lbrace p, H \rbrace\right)$, involving accelerations, is not required to impose restrictions on the gauge parameters. Since this is the only input in our method of abstraction of the independent gauge parameters, we note that our analysis is off-shell.

\subsection{Required algebrae and finding the structure functions}

Before we begin, let us recall that all brackets are computed in the reduced phase space where a sector (second-class) of the original primary constraints has been eliminated by modifying the Poisson algebra. The algebra thus being used was presented in \eqref{algebra1} and its corollary \eqref{algebra2}.

\paragraph*{\it Algebrae within primary first-class sector:} The algebrae in this sector can be calculated directly with the definition of $\pi''{}_i^{\ 0}$ given in \eqref{Constrnt defns}. They all turn out to be either zero, or negligible square of constraint type terms (composed of {\em all} constraints, first and second class).
\begin{align}
\label{algebra PP}
\begin{aligned}
\lbrace \pi''{}_i^{\ 0}, \pi''{}_j^{\ 0} \rbrace &= 0\\
\lbrace \pi''{}_i^{\ 0}, \Pi_j^{\ 0}\rbrace &= 0\\
\lbrace \Pi_i^{\ 0}, \Pi_j^{\ 0} \rbrace &= 0.
\end{aligned}
\end{align}

\paragraph*{\it Algebrae within secondary first-class sector:} This may also be calculated using the basic algebra \eqref{algebra1} and the definitions \eqref{Constrnt defns}. We list these below \cite{Blagojevic:2010ir}
\begin{align}
\label{algebra SS}
\begin{aligned}
\lbrace \bar{\mathcal{H}}_i, \bar{\mathcal{H}}_j \rbrace &= - \epsilon_{ijk} \left( f^{kn}-f\eta^{kn} \right)\\
\lbrace \bar{\mathcal{H}}_i, \bar{\mathcal{K}}_j \rbrace &= - \epsilon_{ijk}\,\bar{\mathcal{H}}^k\\
\lbrace \bar{\mathcal{K}}_i, \bar{\mathcal{K}}_j \rbrace &= - \epsilon_{ijk}\,\bar{\mathcal{K}}^k.
\end{aligned}
\end{align}

\paragraph*{\it Algebrae between primary and secondary first-class:} Note that there are two forms of total hamiltonian; $\mathcal{H}_T$ defined in \eqref{H_T 2} with the Lagrange multipliers for the primary second-class undetermined, and the other $\hat{\mathcal{H}}_T$, with Lagrange multipliers corresponding to the primary second-class fixed \eqref{H_T 3}. These are equal upto terms which are square in constraints  \cite{Blagojevic:2010ir} and hence the difference is ignored. Now we have $\lbrace \mathcal{H}_T, \pi''{}_i^{\ 0} \rbrace = \bar{\mathcal{H}}_i$, and thus
\begin{align}
\lbrace \hat{\mathcal{H}}_T, \pi''{}_i^{\ 0} \rbrace = \bar{\mathcal{H}}_i.
\end{align}
Using the definition of $\mathcal{H}_T$ given in \eqref{H_T 3}, and after performing some manipulations, we arrive at:
\begin{align}
\label{something1}
b^j_{\ 0}\,&\lbrace \bar{\mathcal{H}}_j, \pi''{}_i^{\ 0} \rbrace + \omega^j_{\ 0}\,\lbrace \bar{\mathcal{K}}_j, \pi''{}_i^{\ 0} \rbrace = 0.
\end{align}
Note that the brackets in \eqref{something1} involve the first-class algebra within itself, which is closed. Terms linear in constraints must come from some constraint out of Table \ref{Tab:Constraints}. An inspection of the same table reveals that there exist no combination of constraints such that one multiplied by $b^j_{\ 0}$ cancels out the other multiplied with $\omega^j_{\ 0}$. Thus the brackets in question must themselves be zero. Similarly, the bracket $\lbrace \mathcal{H}_T, \Pi_i^{\ 0} \rbrace = \bar{\mathcal{K}}_i$ results in the other set of brackets (between $\Pi_i^{\ 0}$ and $\bar{\mathcal{H}}_j$ or $\bar{\mathcal{K}}_j$) to also be equal to zero. We list the results below:
\begin{align}
\label{algebra PS}
\begin{aligned}
\lbrace \bar{\mathcal{H}}_j, \pi''{}_i^{\ 0} \rbrace &= 0 \\
\lbrace \bar{\mathcal{K}}_j, \pi''{}_i^{\ 0} \rbrace &= 0 \\
\lbrace \bar{\mathcal{H}}_j, \Pi{}_i^{\ 0} \rbrace &= 0 \\
\lbrace \bar{\mathcal{K}}_j, \Pi{}_i^{\ 0} \rbrace &= 0.
\end{aligned}
\end{align}

\paragraph*{\it Structure functions of the algebrae within first-class:} We can now collect and list the $C^I_{\;\;JK}$'s defined in \eqref{C_defn} from the results of all the previous algebrae calculated in this section. Only the non-vanishing ones are explicitly written.
\begin{align}
\label{Cs}
\begin{aligned}
{\left(C^2_{\;\; 11}\right)}_{ijk}(x'',x,x') &= -\epsilon_{ijn}\left( f^n_{\ k} - f \delta^n_k \right) \,\delta(x-x'') \delta(x''-x')\\
{\left(C^1_{\;\; 12}\right)}_{ijk}(x'',x,x') &= -\epsilon_{ijk} \,\delta(x-x'') \delta(x''-x')\\
{\left(C^2_{\;\; 22}\right)}_{ijk}(x'',x,x') &= -\epsilon_{ijk} \,\delta(x-x'') \delta(x''-x')\\
\end{aligned}
\end{align}
In particular, we see that structure functions for algebrae within the primary first-class vanishes. Also, since the algebrae between any two first-class constraints can be expressed in terms of only the secondary first-class, all the $C^A_{\;\; IJ}$ turn out to be zero.

\paragraph*{\it Algebrae between $\mathcal{H}^{(1)}$ and first-class constraints:} The other other set of required algebrae \eqref{V_defn} can now be calculated using the definition $\mathcal{H}^{(1)}=b^i_{\ 0}\bar{\mathcal{H}}_i + \omega^i_{\ 0}\bar{\mathcal{K}}_i$. We note that this is just a combination of the secondary first-class constraints. So we use the appropriate algebrae between first-class constraints in the calculations.
\begin{align}
\label{algebrae H1_1stCl}
\begin{aligned}
\lbrace \mathcal{H}^{(1)}, \pi''{}_i^{\ 0} \rbrace &= \bar{\mathcal{H}}_i\\
\lbrace \mathcal{H}^{(1)}, \Pi_i^{\ 0} \rbrace &= \bar{\mathcal{K}}_i\\
\lbrace \mathcal{H}^{(1)}, \bar{\mathcal{H}}_i \rbrace &= \epsilon_{ijk} \omega^j_{\ 0} \bar{\mathcal{H}}^k + \epsilon_{ijk} b^j_{\ 0} \left( f^{kn} - f \eta^{kn} \right)\bar{\mathcal{K}}_n\\
\lbrace \mathcal{H}^{(1)}, \bar{\mathcal{K}}_i \rbrace &= \epsilon_{ijk} b^j_{\ 0} \bar{\mathcal{H}}^k + \epsilon_{ijk} \omega^j_{\ 0} \bar{\mathcal{K}}^k
\end{aligned}
\end{align}

\paragraph*{\it Structure functions of $\mathcal{H}^{(1)}$ with first-class sector:} The set $V^I_{\ J}$ defined in \eqref{V_defn} can be read off from the algebrae \eqref{algebrae H1_1stCl} calculated above. We list the non-vanishing ones below:
\begin{align}
\label{Vs}
\begin{aligned}
{\left(V^1_{\;\;\, 1}\right)}_{ik}(x',x) &=  \epsilon_{ijk}\,\omega^j_{\ 0}\,\delta(x-x')\\
{\left(V^2_{\;\;\, 1}\right)}_{ik}(x',x) &=  \epsilon_{ijl} \, b^j_{\ 0} \left( f^l_{\ k} - f \delta^l_k \right)\,\delta(x-x')\\
{\left(V^1_{\;\;\, 2}\right)}_{ik}(x',x) &=  \epsilon_{ijk} \, b^j_{\ 0}\,\delta(x-x')\\
{\left(V^2_{\;\;\, 2}\right)}_{ik}(x',x) &=  \epsilon_{ijk} \,\omega^j_{\ 0}\,\delta(x-x')\\
{\left(V^1_{\;\;\, 3}\right)}_{ik}(x',x) &=  \eta_{ik}\,\delta(x-x')\\
{\left(V^2_{\;\;\, 4}\right)}_{ik}(x',x) &=  \eta_{ik}\,\delta(x-x').\\
\end{aligned}
\end{align}

\subsection{The generator}

Having found all the required structure functions, we can now construct the relations between the gauge parameters $\varepsilon^{(I)}$ given through the master equation \eqref{RB master 2}.
\begin{align}
\label{rel epsilons}
\begin{aligned}
\dot{\varepsilon}^{(1)i} &= \varepsilon^{(3)i} - \varepsilon^{(1)k}\epsilon^i_{\ jk}\omega^j_{\ 0} - \varepsilon^{(2)k}\epsilon^i_{\ jk}b^j_{\ 0}\\
\dot{\varepsilon}^{(2)i} &= \varepsilon^{(4)i} - \varepsilon^{(2)k}\epsilon^i_{\ jk}\omega^j_{\ 0} - \varepsilon^{(1)k}\epsilon_{klj}\left( f^{li} - f\eta^{li} \right) b^j_{\ 0} \\
\end{aligned}
\end{align}
Note that the algebrae between the primary first-class constraints with all other first-class being zero \eqref{Cs}, no $C$-structure function appears in the above relations. After using these equations \eqref{rel epsilons} in the generator \eqref{gen G} to eliminate the gauge parameters $\varepsilon^{(3)}$ and $\varepsilon^{(4)}$, we obtain the generator in terms of the two independent gauge parameters $\varepsilon^{(1)}$ and $\varepsilon^{(2)}$.
\begin{align}
\label{generatorOur}
G = \int d^2x & \left[\left\lbrace \dot{\varepsilon}^{(1)i} +  \varepsilon^{(1)k} \,\epsilon_k^{\ ij}\,\omega_{j0} + \varepsilon^{(2)k} \,\epsilon_k^{\ ij}\,b_{j0}\right\rbrace\,\pi''{}_i^{\ 0} \right.\nonumber\\
&+\left.\left\lbrace \dot{\varepsilon}^{(2)i} + \,\varepsilon^{(2)k} \,\epsilon_k^{\ ij}\,\omega_{j0} + \epsilon^{(1)k} \epsilon_k^{\ lj}\,b_{j0} \left( f_l^{\ i}-f\delta^i_l \right) \right\rbrace \Pi_i^{\ 0} \right.\nonumber\\
&+\left. \varepsilon^{(1)i}\,\bar{\mathcal{H}_i} + \varepsilon^{(2)i}\,\bar{\mathcal{K}_i}\right].
\end{align}
The parameters can be renamed $(\varepsilon^{(1)}=\tau\,,\; \varepsilon^{(2)}=\sigma)$, and the expression \eqref{generatorOur} be arranged to arrive at the generator
\begin{align}
\label{generatorFinal}
\begin{aligned}
G=&\int d^2x \left[\mathcal{G}_\tau(x)+\mathcal{G}_\sigma(x)\right]\\
&\mathcal{G}_\tau=\dot{\tau}^i\,\pi''{}_i^{\ 0} + \tau^i\left[ \bar{\mathcal{H}}_i -\epsilon_{ijk} \omega^j_{\ 0} \pi''{}^{k0} - \epsilon_{ijk} b^j_{\ 0} \left( f^{kn} - f\eta^{kn} \right) \Pi_n^{\ 0} \right]\\
&\mathcal{G}_\sigma=\dot{\sigma}^i\Pi_i^{\ 0} + \sigma^i\left[ \bar{\mathcal{K}}_i - \epsilon_{ijk} \omega^j_{\ 0}\Pi^{k0} - \epsilon_{ijk} b^j_{\ 0} \pi''{}^{k0} \right].
\end{aligned}
\end{align}
The above generator was also reported in \cite{Blagojevic:2010ir}, where an on-shell method of constructing gauge generators following \cite{Castellani:1981us}, was used. Our's however, is an explicitly off-shell method of construction. Also note that that the number of independent gauge parameters here ($3+3=6$) is equal to the total number of (independent) primary first-class constraints (see Table \ref{Tab:Constraints}), as mentioned earlier (see discussion below eq. \ref{RB master 2}).

In the next section, we construct the symmetries of the basic fields obtained by the above generator and study their relation with the underlying Poincar\'{e} symmetries of the model.

\section{The symmetries and their mapping: hamiltonian to Poincar\'{e}}
\label{Sec:Map}

The symmetries of the basic fields $\left( b^i_{\ \mu}, \omega^i_{\ \mu}, f^i_{\ \mu}, \lambda^i_{\ \mu} \right)$ can be calculated using the generator \eqref{generatorFinal} constructed in the previous section. The algebra used \eqref{algebra1} is that defined in the reduced space as explained earlier in Section \ref{Sec:Model}. Thus the variation of the triad field `$b$' is
\begin{align}
\label{delta_G b}
\delta_G b^h_{\ \zeta} = \partial_\zeta \tau^h - \epsilon^h_{\ jk}\tau^j\omega^k_{\ \zeta} - \epsilon^h_{\ jk}\sigma^j b^k_{\ \zeta}.
\end{align}
It is clear from a comparison between this symmetry generated via hamiltonian gauge generator and that of the PGT symmetry \eqref{PGT deltas} of `$b$', that the Poincar\'{e} symmetries of local Lorentz rotation and general diffeomorphism cannot be identified in the set $\delta_G b^h_{\ \zeta}$. We therefore map the arbitrary gauge parameters $\tau^i$ and $\sigma^i$ to the Poincar\'{e} parameters $\xi^\mu$ and $\theta^i$ to recover the Poincar\'{e} symmetries. The map used is
\begin{align}
\label{map}
\begin{aligned}
\tau^i &= -\xi^\rho\,b^i_{\ \rho}\\
\sigma^i &= -\theta^i - \xi^\rho \,\omega^i_{\ \rho}.
\end{aligned}
\end{align}
This type of map was reported earlier in studies of topologically massless \cite{Blagojevic:2004hj,Banerjee:2009vf} as well as massive \cite{Blagojevic:2008bn} models of gravity. In \cite{Banerjee:2009vf}, it was shown explicitly that though both $\delta_G$ and $\delta_{PGT}$ generate off-shell symmetries, they can be related to each other through the map \eqref{map} only on-shell, i.e. upon imposition of the equations of motion. We will shortly see that something similar also happens here.

\noindent On using the above map in \eqref{delta_G b}, we get the following form of symmetry for the triad:
\begin{align}
\label{G PGT B}
\delta_{\scriptscriptstyle G} b^h_{\ \zeta} &= -\partial_\zeta \xi^\rho \, b^h_{\ \rho} - \xi^\rho \,\partial_\zeta b^h_{\ \rho} + \epsilon^h_{\ jk}\xi^\rho b^j_{\ \rho} \omega^k_{\ \zeta} + \epsilon^h_{\ jk}\theta^j b^k_{\ \zeta} + \epsilon^h_{\ jk}\xi^\rho\omega^j_{\ \rho} b^k_{\ \zeta} \nonumber\\
&= -\partial_\zeta \xi^\rho\,b^h_{\ \rho} - \xi^\rho\,\partial_\rho b^h_{\ \zeta} - \epsilon^h_{\ jk}b^j_{\ \zeta}\theta^k + \xi^\rho \left( \partial_\rho b^h_{\ \zeta} - \partial_\zeta b^h_{\ \rho} + \epsilon^h_{\ jk}\omega^j_{\ \rho}b^k_{\ \zeta} - \epsilon^h_{\ jk}\omega^j_{\ \zeta}b^k_{\ \rho} \right) \nonumber\\
&= \delta_{\scriptscriptstyle PGT} b^h_{\ \zeta} + \xi^\rho\, T^h_{\ \rho\zeta}.
\end{align}
We thus recover the PGT symmetry, but modulo terms which vanish on-shell. To see this, note the equation of motion \eqref{EOM lambda} corresponding to the field $\lambda$. Since torsion is antisymmetric in its Greek indices, i.e. $T^i_{\ \mu\nu} = -T^i_{\ \nu\mu}$, we have $$\epsilon^{\mu\nu\rho}\,T^i_{\ \nu\rho} = 0 \quad \Rightarrow \quad T^i_{\ \nu\rho} = 0.$$ This phenomenon -- that among all the equations of motion the imposition of vanishing torsion is required to come back to the PGT (local Lorentz + diffeomorphisms) symmetries, from the hamiltonian gauge symmetries, is remarkable. As was earlier noted, the difference between the original and PGT formulation of BHT theory lies in that the former is built on Reimannian spacetime (only curvature, zero torsion), while the latter on Riemann-Cartan spacetime (both curvature and torsion). So we do not find it surprising that the triad field $b^i_{\ \mu}$, which alone makes up the metric $g_{\mu\nu}$, is restored to its expected PGT symmetry by use of zero torsion condition.

Coming back to hamiltonian gauge symmetries, let us examine another field, the axillary field `$f$'. The gauge transformation of `$f$' generated by the generator \eqref{generatorFinal} is
\begin{align}
\label{delta_G f}
\delta_{\scriptscriptstyle G} f^h_{\ \zeta} &= \dot{\tau}^i f_i^{\ h}\,\delta^0_\zeta + \partial_\alpha\left( \tau^i f^h_{\ i} \right)\delta^\alpha_\zeta - \epsilon^h_{\ jk}\tau^i f^j_{\ i}\omega^k_{\ \alpha}\delta^\alpha_\zeta + \tau^i b_i^{\ \mu} \left( \nabla_\mu f^h_{\ k} \right)b^k_{\ 0}\delta^0_\zeta - \epsilon_{ijk} \tau^i \omega^j_{\ 0} f^{kh} \delta^0_\zeta \nonumber \\
&\quad - \left( \frac{m^2}{a} \right) \epsilon^h_{\ jk}\tau^j\lambda^k_{\ \alpha} \delta^\alpha_\zeta - \left( \frac{m^2}{a} \right) \epsilon^h_{\ jk} \tau^i \lambda^j_{\ i} b^k_{\ \alpha} \delta^\alpha_\zeta - \epsilon^h_{\ jk}\sigma^j f^k_{\ \zeta}.
\end{align}
Use of the map \eqref{map} in the above transformation gives
\begin{align}
\label{delta_G f mapped PGT}
\delta_{\scriptscriptstyle G} f^h_{\ \zeta} = \delta_{\scriptscriptstyle PGT} f^h_{\ \zeta} &+ \frac{m^2}{a} ~\epsilon_{\mu\rho\zeta}\,\epsilon^{\mu\nu\sigma} ~\xi^\rho \left[ \frac{a}{m^2} \nabla_\nu f^h_{\ \sigma} + \epsilon^h_{\ jk}b^j_{\ \nu}\lambda^k_{\ \sigma} \right] \nonumber\\
&+ \frac{m^2}{a} ~\epsilon_{\mu 0 \rho}\,\epsilon^{\mu 0 \sigma} ~\xi^\rho\left[ \frac{a}{m^2} \nabla_0 f^h_{\ \sigma} + \epsilon^h_{\ jk}b^j_{\ 0}\lambda^k_{\ \sigma} \right]\delta^0_\zeta - \xi^\rho\,T^i_{\ 0\rho}\,f^h_{\ i}\,\delta^0_\zeta\,.
\end{align}
As seen earlier in case of the triad, here too we see that the hamiltonian symmetry is equal to the PGT symmetries modulo the equations of motion. In this case, the equations of motion for the fields `$\lambda$' and `$\omega$' \eqref{EOM} are required to identify with the PGT symmetries. Also in the above computations, use of the constraint $\Theta^{ij}=f^{ij}-f^{ji}$ is required.

The symmetries for the other two fields `$\omega$' and `$\lambda$' also give similar results, only the algebraic nature is more involved. Thus all the fields have two sets of symmetries, the PGT symmetries and the hamiltonian gauge symmetries. Both of these are off-shell in nature. But they can be identified with each other only on-shell.

Now, a subtle issue arises in this identification of the two symmetries through the use of the map \eqref{map}. The map between the two sets of independent gauge parameters is field dependent in nature. As a result, one may wonder whether one can use this map at the level of the generator, i.e. before computation of symmetries. Once the map is used in the generator itself, it will give rise to non-trivial brackets with other fields when computing the symmetries. The proper way to frame the question would be to study the commutativity manifested in the diagram:
\begin{center}
$\begin{CD}
G[\tau,\sigma] @>>> \delta_{[\tau,\sigma]} @.\\
@VV\text{Map}V @VV\text{Map}V @.\\
G[\xi,\theta] @>>> \delta_{[\xi,\theta]} @>>\text{\phantom{www}on-shell\phantom{www}}> \delta_{PGT}
\end{CD}$
\end{center}
The issue however can be resolved on noting that the generator is nothing but a combination of (first-class) constraints multiplied by the gauge parameters, and possibly, other fields. Thus, when terms apart from the constraints (in this case - especially the parameters) gives rise to brackets, they are rendered insignificant due to multiplication with a constraint. So it is immaterial whether we use the map in the gauge symmetries generated by the hamiltonian generator, or in the mapped generator itself.

If we use the map \eqref{map} in the generator \eqref{generatorFinal} constructed in the previous section, we get, upto terms proportional to square of constraints
\begin{align}
\label{generator mapped}
G = &-\dot{\xi}^\mu \left[ b^i_{\ \mu}\pi_i^{\ 0} + \omega^i_{\ \mu}\Pi_i^{\ 0} + \lambda^i_{\ \mu}p_i^{\ 0} + f^i_{\ \mu}P_i^{\ 0} \right] - \xi^\mu \left[ b^i_{\ \mu}\hat{\mathcal{H}}_i + \omega^i_{\ \mu}\mathcal{K}_i + \lambda^i_{\ \mu}\mathcal{T}_i + f^i_{\ \mu}\hat{\mathcal{R}}_i \right.\nonumber\\
&\left. \qquad\qquad\qquad\qquad\qquad\quad + ~(\partial_\mu b^i_{\ 0})\pi_i^{\ 0} + (\partial_\mu \omega^i_{\ 0})\Pi_i^{\ 0} + (\partial_\mu \lambda^i_{\ 0})p_i^{\ 0} + (\partial_\mu f^i_{\ 0})P_i^{\ 0} \right]\nonumber \\
&- \dot{\theta}^i \Pi_i^{\ 0} - \theta^i\left[ \mathcal{K}_i -\epsilon_{ijk}\left( b^j_{\ 0}\pi^{k0} + \omega^j_{\ 0}\Pi^{k0} + \lambda^j_{\ 0}p^{k0} + f^j_{\ 0}P^{k0} \right) \right].
\end{align}
This generator also generates symmetries of the basic fields and these agree with those of PGT {\em on-shell} \cite{Blagojevic:2010ir}. Thus our results are in agreement with the above conclusion of commutativity of the diagram given above.

\subsection{Consistency check}

We will now finally show an internal consistency check of the algorithm given through the relation \eqref{RB master 1} obtained in section \ref{Sec:Gen}. This relation, unlike its twin \eqref{RB master 2}, is not a new restriction on the gauge parameters as it can be independently derived through use of \eqref{RB master 2} and the generator \eqref{generatorFinal} as was shown in \cite{Banerjee:1999hu}. Note that the construction of the generator itself is independent of \eqref{RB master 1}. We start with an observation on the equation of motion of the field $b^i_{\ 0}$
\begin{align}
\label{something2}
\dot{b}^i_{\ 0} = \lbrace b^i_{\ 0}, \int \mathcal{H}_T \rbrace = u^i_{\ 0},
\end{align}
where in the last step, we used the total hamiltonian density given in \eqref{H_T 3}. The variation of the Lagrange multiplier $u^i_{\ 0}$ can thus be obtained from the variation of the field $b^i_{\ 0}$ calculated in \eqref{delta_G b}
\begin{align}
\label{something3}
\delta b^i_{\ 0} = \partial_0 \tau^i - \epsilon^i_{\ jk} \tau^j \omega^k_{\ 0} - \epsilon^i_{\ jk}\sigma^j b^k_{\ 0}.
\end{align}
So, we have
\begin{align}
\label{something4}
\delta u^i_{\ 0} = \frac{d}{dt} \delta b^i_{\ 0} = \frac{d}{dt}\left[ \dot{\tau}^i - \epsilon^i_{\ jk}\tau^j\omega^k_{\ 0} - \epsilon^i_{\ jk}\sigma^j b^k_{\ 0} \right] = \dot{\varepsilon}^{(3)}{}^i.
\end{align}
Use has been made of the redefinitions $(\varepsilon^{(1)}=\tau\,,\; \varepsilon^{(2)}=\sigma)$ and the relations between the gauge parameters \eqref{rel epsilons} which were found by employing only the second relation \eqref{RB master 2}.
Turning now to the first relation \eqref{RB master 1} that also gives variations of the Lagrange multipliers, we see for $A=3$, i.e. $\chi^{(3)}=u^i_{\ 0}$,
\begin{align}
\label{something5}
\delta \chi^{(3)} = \delta u^i_{\ 0} = \frac{d \varepsilon^{(3)}{}^i}{dt} - \int d^2x ~\varepsilon^{(I)}{}^k (V^3_{\; I}){}^i_{\ k } - \int d^2x ~\varepsilon^{(I)}{}^k \int d^2x' ~\chi^{(B)}{}^j (C^3_{\;\; IB}){}^i_{\ jk}.
\end{align}
Since $V^3_{\; I}=0$ \eqref{Vs} and $C^3_{\;\; IB}=0$ \eqref{Cs}, we finally get
\begin{align}
\label{something6}
\delta u^i_{\ 0} = \frac{d \varepsilon^{(3)}{}^i}{dt},
\end{align}
which is nothing but \eqref{something4}. This shows the internal consistency of our scheme.

\section{Discussions}
\label{Sec:Disc}

In this paper we have constructed the hamiltonian gauge generator of the cosmological BHT model described in first-order form by the action \eqref{lagrangian} using a completely off-shell method. We have explicitly found the hamiltonian gauge symmetries resulting from this generator and shown that these symmetries can be mapped to the Poincar\'{e} symmetries only on-shell, through a mapping of the gauge parameters. Remarkably the vanishing torsion condition, which takes us from the Riemann-Cartan spacetime of first order PGT formulation to the usual metric formulation in Riemann spacetime, plays an important role in this on-shell mapping. We also noted that the map used by us, which is also quite common in the literature \cite{Banerjee:2009vf,Blagojevic:2004hj,Blagojevic:2008bn}, is field dependent. We clarify why this does not cause any problem in computation of the symmetries through the generator. It can be used both directly in the generator, i.e. before computation of symmetries and also after computation of symmetries through the generator. The two processes were shown to be equivalent.

Finally we would like to comment that our results would be useful in finding the corresponding conserved charges of BHT gravity consistently at an off-shell level. This would in turn play an important role in the obtention of the central charges of the asymptotic symmetry.


\end{document}